\documentclass[aps,twocolumn,superscriptaddress,preprintnumbers,showpacs]{revtex4}
\usepackage{epsfig}
\usepackage{graphicx,amssymb,epsfig,amsmath}
\begin{document}

\title{Ferromagnetic resonance study of Co/Pd/Co/Ni multilayers with perpendicular anisotropy irradiated with Helium ions}
\author{J-M. L. Beaujour, A. D. Kent} 
\affiliation{Department of Physics, New York University, 4
Washington Place, New York, NY 10003, USA}
\author{D. Ravelosona}
\affiliation{Institut d'Electronique Fondamentale, UMR CNRS 8622, Universite Paris Sud, 91405 Orsay Cedex,
France}
\author{I. Tudosa and E. E. Fullerton}
\affiliation{University of California, San Diego, Center for Magnetic Recording Research, La Jolla, CA 92093-0401, USA}
\date{\today}

\begin{abstract}
We present a ferromagnetic resonance (FMR) study of the effect of Helium ion irradiation on the magnetic anisotropy, the linewidth and the Gilbert damping of a Co/Ni multilayer coupled to Co/Pd bilayers. The perpendicular magnetic anisotropy decreases linearly with He ion fluence, leading to a transition to in-plane magnetization at a critical fluence of $5 \times 10^{14}$ ions/cm$^2$. We find that the damping is nearly independent of fluence but the FMR linewidth at fixed frequency has a maximum near the critical fluence, indicating that the inhomogeneous broadening of the FMR line is a non-monotonic function of the He ion fluence. Based on an analysis of the angular dependence of the FMR linewidth, the inhomogeneous broadening is associated with spatial variations in the magnitude of the perpendicular magnetic anisotropy. These results demonstrate that ion irradiation may be used to systematically modify the magnetic anisotropy and distribution of magnetic anisotropy parameters of Co/Pd/Co/Ni multilayers for applications and basic physics studies.
\end{abstract}
\maketitle

Co/Ni multilayers are of great interest in information technology and in spin-transfer devices because they combine high spin polarization with large perpendicular magnetic anisotropy (PMA) \cite{Mangin2006,Mangin2009,Bedau2010}. Perpendicular anisotropy was predicted in Co/Ni multilayers and has been shown experimentally to be a function of layer composition and thin film growth conditions \cite{Daalderop1992}. Recently it has been shown that the coercivity and perpendicular anisotropy of a multilayer can be tailored by Helium ion irradiation \cite{Chappert1998}, making it possible to modify films after growth to tune their magnetic properties. This is of great interest for applications and basic physics studies, as in many cases the perpendicular anisotropy of a structure sets importance device metrics, and ion irradiation offers the possibility of changing these properties, both locally and globally, after device fabrication. For instance, in spin-transfer magnetic random access memories (STT-MRAM) the current threshold for switching is proportional to the PMA \cite{Sun2000,Mangin2009}.

Light ion irradiation has been used to vary the magnetic properties of multilayer films in many earlier studies \cite{Traverse1989,Devolder2000,Fassbender2004,Bilzer2008,Stanescu2008}. For instance, the coercivity of Co/Pt multilayers was found to decrease with ion dose  \cite{Devolder2000}. This behavior was attributed to interface mixing and strain relaxation reducing the PMA. Very recently, it was reported that the coercive field of Co/Ni multilayers decreases linearly with increasing He$^+$ irradiation fluence up to \begin{sffamily}\textit{F} \end{sffamily}$=10^{15}$ ions/cm$^2$, suggesting changes in the magnetic anisotropy of the film \cite{Stanescu2008}. The effect of ion irradiation on the FMR linewidth has also been studied in Au/Fe multilayer films with PMA \cite{Bilzer2008}. The PMA is reduced by He$^+$ irradiation and the authors explained this by a reduction of the inhomogeneous contribution to the FMR linewidth. In a recent paper, we presented a FMR study of the anisotropy and the linewidth of a Co/Ni multilayer film exposed to a relatively high He$^+$ irradiation fluence ( \begin{sffamily}\textit{F} \end{sffamily}$=10^{15}$ ions/cm$^2$) \cite{Beaujour2009}. In addition to a strong decrease of the PMA, the contribution to the linewidth from spatial variation of the anisotropy, was reduced compared to that of a non irradiated Co/Ni multilayer. Furthermore, a correlation between the anisotropy distribution and the linewidth broadening from two-magnon scattering (TMS) mechanism was observed. However, a systematic study of the effect of the He$^+$ irradiation on the FMR spectra as a function of fluence has yet to be reported.

In this paper, we present a FMR study of a Co/Ni multilayer coupled to Co/Pd bilayers exposed to Helium ion irradiation of fluence up to $10^{15}$ ions/cm$^2$. The PMA and the contributions to the FMR linewidth, including those from Gilbert damping ($\alpha$), are studied as a function of fluence.

The samples had the following layer structure: $||$3 Ta$|$1 Pd$|$0.3 Co$|$1 Pd$|$0.14 Co$|$[0.8 Ni$|$0.14 Co]$\times 3|$1 Pd$|$0.3 Co$|$1 Pd$|$3 Ta$||$ (layer thickness in nm) and was fabricated by dc magnetron sputtering. The Co/Ni multilayer is embedded between Co/Pd bilayers to enhance the overall PMA of the film and to have resonance frequencies in which the full angular dependence of the FMR response could be investigated in a $1$ T electromagnet. The substrate was cleaved into several pieces that were then exposed to different doses of Helium ion irradiation of energy 20 keV with fluence in the range $10^{14} \leq $ \begin{sffamily}\textit{F} \end{sffamily} $\leq 10^{15}$ ions/cm$^2$. FMR measurements were conducted at room temperature using a coplanar waveguide (CPW). Details of the experimental setup can be found in \cite{Beaujour2007}. The field swept CPW transmission signal ($S_{21}$) was recorded as a function of frequency for dc magnetic fields normal to the film plane and as a function of the out-of-plane field angle at 20 GHz. The magnetization density of the film at room temperature was measured with a SQUID magnetometer: $M_{\mathrm{s}} \simeq 4.75 \times 10^5$ A/m.  Within the measurement uncertainty, $M_{\mathrm{s}}$ remains unchanged after irradiation. 

\begin{figure}
\begin{center}\includegraphics[width=8.5cm]{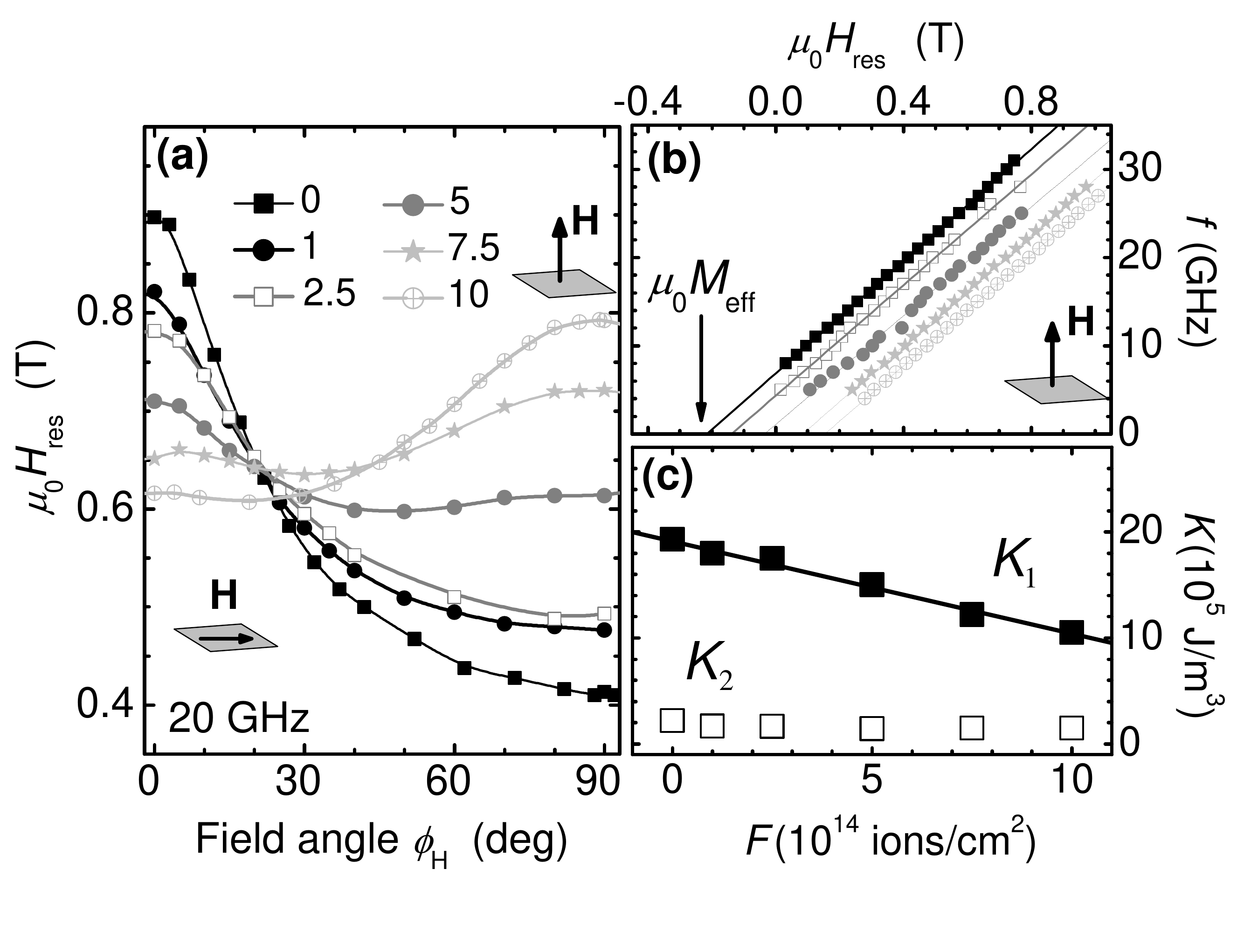}
\vspace{-6 mm}\caption{\label{fig1} a) Angular dependence of the resonance field for different irradiation fluence ($\times 10^{14}$ ions/cm$^2$).  The solid lines are guides to the eye. b) Frequency dependence of the resonance field when the applied field is normal to the film plane ($\phi_{H}=90^{\mathrm{o}}$) at selected fluences. The solid lines are the fits to Eq.~\ref{ResCond}. The zero frequency intercept gives the effective demagnetization field, $\mu_0 M_{\mathrm{eff}}$. c) The second and fourth order perpendicular anisotropy constants, $K_1$ and $K_2$, versus fluence.}
\end{center}
\end{figure}

Fig. \ref{fig1}a shows the out of plane angular dependence of the resonance field at 20 GHz for different fluences. For the non-irradiated film, the resonance field when $\bf{H}$ is normal to the film plane ($\phi_{\mathrm{H}}=90^{\mathrm{o}}$) is smaller than that when the field is in the film plane ($\phi_{\mathrm{H}}=0^{\mathrm{o}}$). This shows that the magnetic easy axis is normal to the film plane. As the fluence increases, $H_{\mathrm{res}}$ at $\phi_H=90^{\mathrm{o}}$ increases whereas that at $\phi_H=0^{\mathrm{o}}$ decreases. In the high fluence range, \begin{sffamily}\textit{F} \end{sffamily} $> 5 \times 10^{14}$ ions/cm$^2$, $H_{\mathrm{res}}$ is the larger when the field is normal to the film plane, i.e. the magnetic easy axis is in the film plane. Fig. \ref{fig1}b shows the frequency dependence of the resonance field for different fluences when the dc field is normal to the film plane. This data is fitted to the resonance condition \cite{Vonsovski1966}:
\begin{equation}
f=\frac{1}{2\pi}\gamma \mu_0(H_{\mathrm{res}}-M_{\mathrm{eff}}),
\label{ResCond}
\end{equation}
where $\gamma$ is the gyromagnetic ratio. $\mu_0 M_{\mathrm{eff}}$, the effective easy plane anisotropy, is given by: $M_{\mathrm{eff}}=M_{\mathrm{s}}-2K_1/(\mu_0 M_{\mathrm{s}})$, where $K_1$ is the second order anisotropy constant. We find that $\mu_0 M_{\mathrm{eff}}$ is negative at low fluence which implies that the PMA is sufficient to overcome the demagnetizing energy and hence the easy axis is normal to the film plane. As the fluence is further increased, $\mu_0 M_{\mathrm{eff}}$ becomes positive. These results confirm that there is a re-orientation of the easy axis, as was inferred indirectly through magnetic hysteresis loop measurements in Ref.~\cite{Stanescu2008}.  $\mu_0 M_{\mathrm{eff}}$ changes sign for fluence between 5 and 7.5 $\times 10^{14}$ ions/cm$^2$. Therefore, by exposing the film to a specific fluence, it is posible to engineer the anisotropy so that the PMA field just compensates the demagnetization field. 

The second order perpendicular anisotropy constant $K_1$ decreases linearly with fluence (Fig. \ref{fig1}c). The film irradiated at 10$^{15}$ ions/cm$^2$ has an anisotropy constant 40\% smaller than that of the non-irradiated film. The 4$^{\mathrm{th}}$ order anisotropy constant $K_2$ is determined from the angular dependence of $H_{\mathrm{res}}$ for magnetization angles $45 \leq \phi \leq 90^{\mathrm{o}}$ \cite{Beaujour2007}. $K_2$ is smaller than $K_1$ by a factor 10, and is nearly independent of fluence.
\begin{figure}
\begin{center}
\includegraphics[width=8.5cm]{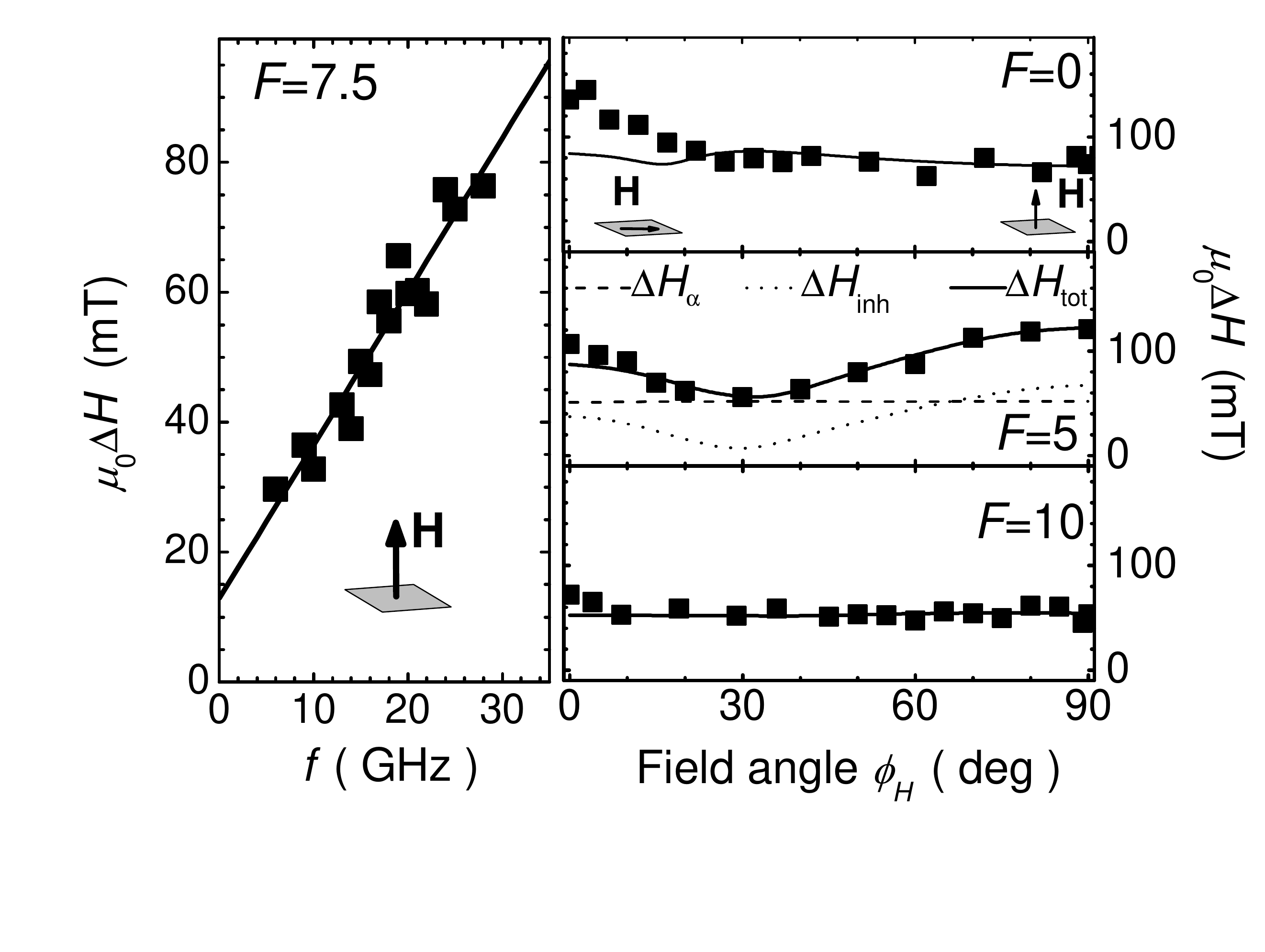}
\vspace{-8 mm}\caption{\label{fig2} On the left, the linewidth as a function of frequency for the film irradiated at 7.5$\times 10^{14}$ ions/cm$^2$. The solid line is a linear fit to the experimental data. On the right, the angular dependence of the linewidth at 20 GHz for a selection of fluences. The solid lines represent the fits to the total linewidth $\Delta H_{\mathrm{tot}}=\Delta H_{\alpha}+\Delta H_{\mathrm{inh}}$, , where the intrinsic damping and the inhomogeneous contribution are represented by the dashed line and the dotted line respectively.}
\end{center}
\end{figure}

The FMR linewidth $\mu_0 \Delta H$ when the dc field is applied normal the film plane was measured as a function of frequency. Fig. \ref{fig2} shows $\mu_0 \Delta H$ versus $f$ for the film irradiated at  \begin{sffamily}\textit{F} \end{sffamily}$=7.5 \times 10^{14}$ ions/cm$^2$. The linewidth increases linearly with frequency, characteristic of Gilbert damping, an intrinsic contribution to the linewidth $\Delta H_\alpha$ \cite{Hillbrandsbook}:
\begin{equation}
\Delta H_\alpha= \frac{4 \pi \alpha}{\mu_0\gamma} f.
\label{Linewidth}
\end{equation}
From a linear fit to the experimental data, the magnetic damping constant is estimated from the slope of the line: $\alpha=0.037 \pm 0.004$. The films irradiated at  \begin{sffamily}\textit{F} \end{sffamily}$=0,$ 1 and 10 $\times 10^{14}$ ions/cm$^2$ shows a similar frequency dependence of the linewidth and have about the same damping constant, $\alpha \approx 0.04$. At intermediate fluence  \begin{sffamily}\textit{F} \end{sffamily}$=2.5$, $5 \times 10^{14}$ ions/cm$^2$, the linewidth is enhanced and is frequency independent, i.e. the linewidth is dominated by an inhomogeneous contribution, $\Delta H_{\mathrm{inh}}$. The angular dependence of the linewidth measured at 20 GHz is shown in Fig. \ref{fig2} for films irradiated at selected fluences. For the non-irradiated film and the film irradiated at 10$^{15}$ ions/cm$^2$, the linewidth is practically independent of the field angle from about $30^{\mathrm{o}}$ up to $90^{\mathrm{o}}$. For the film irradiated at $5\times 10^{14}$ ions/cm$^2$, $\Delta H$ is clearly angular dependent and shows a minimum at an intermediate field angle. 

The angular dependence of the linewidth was fit to a sum of the intrinsic linewidth $\Delta H_{\alpha}$ and an inhomogeneous contribution $\Delta H_{\mathrm{inh}}$ for magnetization angles $45^{\mathrm{o}}\leq \phi \leq 90^{\mathrm{o}}$, an angular range in which TMS does not contribute to the linewidth \cite{Beaujour2007}. The inhomogeneous linewidth is given by:
\begin{equation}
\Delta H_{\mathrm{inh.}} (\phi_H)= \left|\partial H_{\mathrm{res}}/\partial K_1\right| \Delta K_1+\left|\partial H_{\mathrm{res}}/\partial \phi \right| \Delta \phi,
\label{Inhomogeneous}
\end{equation}
where $\Delta K_1$ is the width of the distribution of anisotropies and $\Delta \phi$ is the distribution of the angles of the magnetic easy axis relative to the film normal. The computed linewidth contributions are shown for the film irradiated at  \begin{sffamily}\textit{F} \end{sffamily}$=5\times 10^{14}$ ions/cm$^2$ in Fig. \ref{fig2}. Note that the intrinsic contribution $\Delta H_{\alpha}$ is practically independent of field angles, as expected when the angle between the magnetization and the applied field is small. For this sample, the maximum angle is about 5$^\mathrm{o}$ and it is due to the fact that the resonance field ($H_{\mathrm{res}}\simeq 0.6$ T) is much larger than the effective demagnetization field ($M_{\mathrm{eff}}\simeq 0$). The inhomogeneous contribution from the distribution in the anisotropy field directions does not significantly affect the fit. For the film irradiated at the lower and upper fluence range, the angular dependence of the intrinsic linewidth is computed fixing the value of $\alpha$ to that obtained from the fit of the frequency dependence of the linewidth. For the other films ( \begin{sffamily}\textit{F} \end{sffamily}=2.5 and 5 $\times 10^{14}$ ions/cm$^2$), $\alpha$ was a fitting parameter.

\begin{figure}
\begin{center}
\includegraphics[width=7cm]{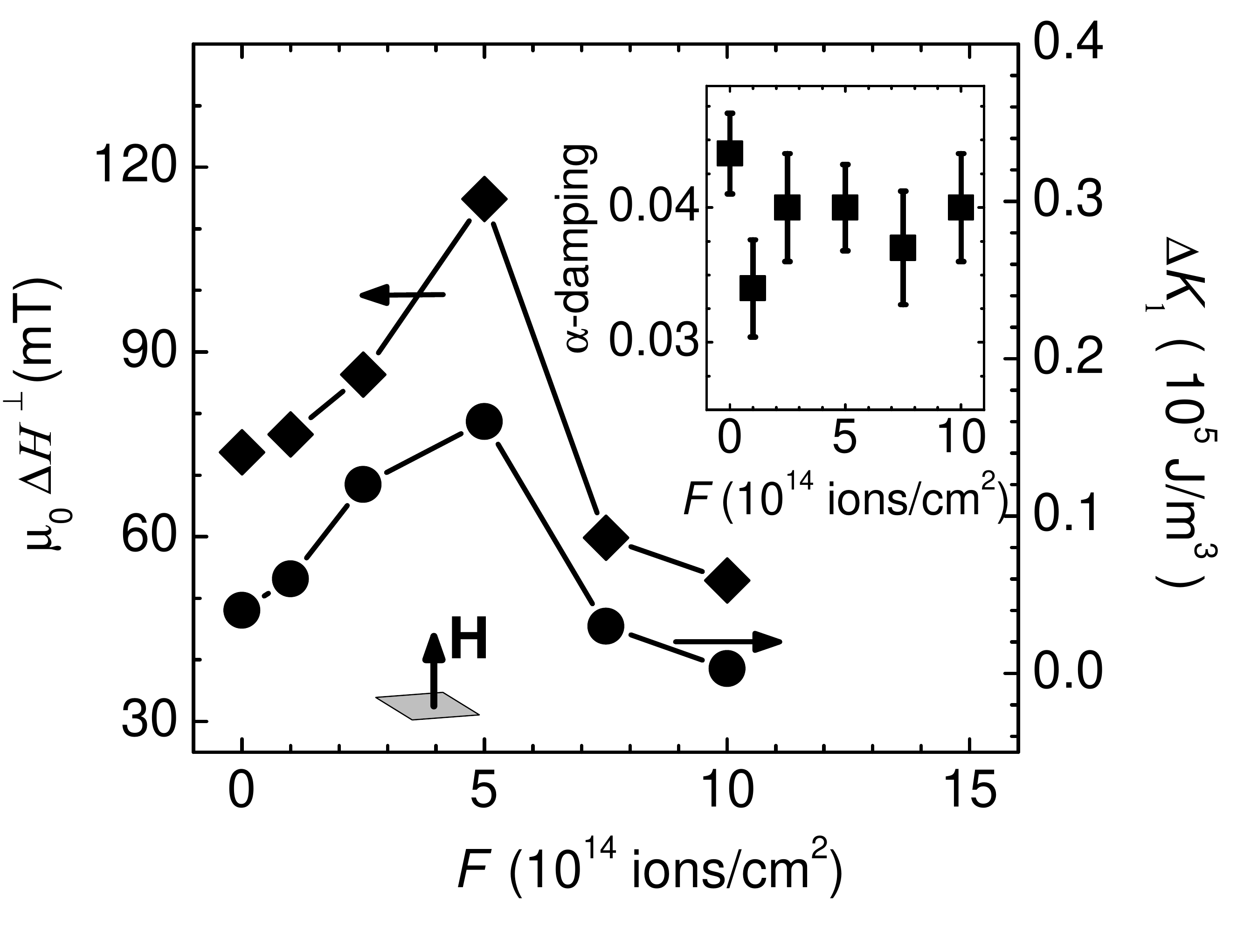}
\vspace{-4 mm}\caption{\label{fig3} The fluence dependence of the linewidth at 20 GHz when the dc field is normal to the film plane (squares). The solid circles represent the fluence dependence of the distribution in the PMA constant $K_1$ determined from fitting $\Delta H$ vs. $\phi_H$. The inset shows the Gilbert damping constant $\alpha$ as a function of fluence.}
\end{center}
\end{figure}

The fluence dependence of $\Delta K_1$ and the linewidth at 20 GHz are shown in Fig. \ref{fig3}. The inset shows the Gilbert damping constant as a function of fluence.
The linewidth at 20 GHz when the field is normal to the film plane is a non monotonic function of fluence. $\Delta H$ increases as the fluence increases, reaching a maximum value at  \begin{sffamily}\textit{F} \end{sffamily}$ \approx 5 \times 10^{14}$ ions/cm$^2$. Then, as the fluence is further increased, $\Delta H$ decreases and falls slightly below the range of values at the lower fluence range. Interestingly, the larger linewidth is observed just at the fluence for which $\mu_0 M_{\mathrm{eff}}=0$. The magnetic damping is practically not affected by irradiation within the error bars: $\alpha \approx 0.04$. The distribution of PMA constants, $\Delta K_1$, shows a similar fluence dependence as the total linewidth, with a maximum at  \begin{sffamily}\textit{F} \end{sffamily}$ \approx 5 \times 10^{14}$ ions/cm$^2$, clearly indicating that this is at the origin of the fluence dependence of the measured linewidth. The distribution in PMA anisotropy is almost zero when the fluence is above $7\times 10^{14}$ ions/cm$^2$. The largest value of $\Delta K_1$ corresponds to variation of $K_1$ ofabout  8\%, which is much larger than that of non irradiated film and the highly irradiated film, $\Delta K_1 / K_1 \approx 2\%$ and 0.3\% respectively.

In summary, irradiation of Co/Pd/Co/Ni films with Helium ions leads to clear changes in its magnetic characteristics, a significant decrease in magnetic anisotropy and a change in the distribution of magnetic anisotropies. Importantly, this is achieved without affecting the film magnetization density and magnetic damping, which remain virtually unchanged. It would be of interest to have a better understanding of the origin of the maximum in the distribution of magnetic anisotropy at the critical fluence, the fluence needed to produce a reorientation of the magnetic easy axis. Nonetheless, these results clearly demonstrate that ion irradiation may be used to systematically tailor the magnetic properties of Co/Pd/Co/Ni multilayers for applications and basic physics studies.

\end{document}